\documentclass[aps,prd,preprint]{revtex4-1}

\usepackage{epsfig}

\begin{document}

\title{Bag model and quark star}
\author{Hua Li$^{1}$, Xin-Lian Luo$^{2}$, and Hong-Shi Zong$^{1,3}$}
\address{$^{1}$ Department of Physics, Nanjing University, Nanjing 210093, China}
\address{$^{2}$ Department of Astronomy, Nanjing University, Nanjing 210093, China}
\address{$^{3}$ Joint Center for Particle, Nuclear Physics and Cosmology, Nanjing 210093, China}

\newcommand{\pslash}{p\!\!\!/\,}
\newcommand{\qslash}{q\!\!\!/\,}
\newcommand{\kslash}{k\!\!\!/\,}
\newcommand{\Pslash}{P\!\!\!\!/\,}

\begin{abstract}
In this paper, incorporating the property of the vacuum negative pressure, 
namely, the bag constant, we presented a new model of the equation of
state (EOS) of quark matter at finite chemical potential and zero
temperature. By comparing our EOS with Fraga {\it et~al.}'s EOS and SQM1 model, one find that
our EOS is softer than Fraga {\it et~al.}'s EOS and SQM1 model. The reason for this difference is analyzed. With these results we investigate the structure of quark star. A
comparison between our model of quark star and other models is made.
The obtained mass of quark star is $ 1.3 \sim 1.66 M_\odot$ and the radius is
$9.5 \sim 14 Km$. One can see that our star's compactness is smaller than
that of other two models. 

\bigskip

\bigskip


\end{abstract}

\maketitle

The research of compact objects is always highly concerned by physicists and astrophysicists \cite{shapiro,Glendenning}, because only these objects can test the fundamental principles of physics under extreme conditions, for example, high temperature and high density circumstance in the core of compact objects. After the conception of quark has been proposed by Gell-Mann and Zweig \cite{Gell-Mann,Zweig}, some authors pointed out that neutron star may be composed of quark matter \cite{Ivanenko,Itoh,Iwamoto,Bodmer,Witten,Haensel,Alcock}.
In Ref. \cite{Witten}, Witten conjectured that strange quark matter may be the true ground state of strong interaction. Inspired by this point of view, some authors raised the conception of quark star. In order to investigate the structure and property of this kind of star, people have developed various models. The equation of state (EOS) of quark matter plays a crucial role for determining the star's structure at high density (above the nuclear density: $2.4*10^{14}g/cm^3$) and high temperature ($T\sim 10~MeV$). The theoretical foundation of all these models is Quantum Chromodynamics (QCD). However, at present it is not possible to obtain reliably the exact EOS of quark matter from first principles of QCD. So, people try to find approximate methods which incorporate basic features of QCD. For example, MIT bag model \cite{Weisskopf,Weber,Soff,Paris}, NJL model \cite{Rehberg,Greiner,Ruster,Menezes}, and perturbative QCD model \cite{Freedman,Baluni,Fraga,Farhi}. These models incorporate some basic features of QCD. For example, the MIT bag model provide mechanism of quark confinement, NJL model can describe dynamical chiral symmetry breaking of QCD, while perturbative QCD model work well at high energy scale due to asymptotic freedom. But all these models have their own weaknesses. In MIT bag model, quarks in the bag are considered as a free Fermi gas. Obviously, MIT bag model violates chiral symmetry even in the limit of massless quark. So, some authors have presented some modified bag models, for example, chiral bag model \cite{Thomas,Hosaka}, which uses pion-quark coupling mechanism to restore chiral symmetry. The NJL model assumes that the interaction between quarks is point-like, so this model is not renormalizable, and it cannot incorporate quark confinement \cite{Klevansky}. Perturbative QCD cannot be applied to low energy region, and cannot describe chiral phase transition and chiral symmetry breaking. From these one can see how difficult it is to understand the property of the compact matter. But by the endeavor mentioned above, people have learned much valuable information about them. Then it deserves to develop some phenomenal methods for studying compact matter. In this paper, we propose a renormalizable EOS based on path integral formalism of QCD and incorporate bag model, then apply it to investigate the structure of
quark star, and compare it with the models mentioned above.

As is well known, if we know the thermodynamic potential of a system, then all thermodynamic variables can be determined \cite{Kapusta}. So, using various methods to get the system's thermodynamic potential is very important. In Refs. \cite{Sun,Roberts,Schmidt,Roberts1,Zong,Feng,He,XY,Y}, one developed a series of nonperturbative methods to demonstrate the aspects of strong interaction at finite density and zero
temperature. Especially, by means of path integral formalism and employing the quark propagator proposed in Ref. \cite{Maris}, a new EOS of QCD at zero temperature and finite chemical potential has been proposed by some of the same authors in Ref. \cite{zong2}. The
purpose of this paper is to incorporate the property of the vacuum negative pressure, 
namely, the bag constant into this EOS of QCD to give a new model for studying the quark star.
In order to be self-contained, here let us first give a brief introduction to the EOS of QCD
proposed in Ref. \cite{zong2}. According to Ref. \cite{zong2}, the quark number density reads:
\begin{equation}
\label{NumDens}
\rho(\mu) = -N_c N_f Z_2\int\frac{d^4p}{(2\pi)^4} tr\left[G_r(\mu)(p)\gamma_4\right],
\end{equation}
where $G_r(\mu)(p)$ is the renormalized quark propagator at finite chemical potential $\mu$, $Z_2$ is the quark wave-function renormalization constant, $N_c$ and $N_f$ denote the number of colours and of flavors, respectively, and the trace operation is over Dirac indices. The model quark propagator chosen in Ref. \cite{zong2} is
\begin{equation}
\label{propagator}
G_r(\mu)(p)=Z_2^{-1}(\zeta^2,\Lambda^2)\sum_{j=1}^{n_p}\left(\frac{r_j}
{i\widetilde{\pslash}+m_j}+\frac{r_j}{i\widetilde{\pslash}+m_j^{*}}\right),
\end{equation}
where ${\widetilde p}=({\vec p}, p_4+i\mu)$, $m_j=a_j+i b_j$ are complex mass scales. If one sets $\mu=0$ in Eq. (\ref{propagator}), one will obtain the quark propagator at zero chemical potential, which is proposed by the authors of Ref. \cite{Maris} under the guidance of the solution of the coupled set of DSEs for the ghost, gluon and quark propagator in Landau gauge. The propagator of this form has $n_P$ pairs of complex conjugate
poles located at $a_j\pm ib_j$. When some $b_j$ is set to zero, the
pair of complex conjugate poles degenerates to a real pole. The
residues $r_j$ are real (note that a similar meromorphic form of the quark propagator was previously proposed in Ref. \cite{Tandy}, in which the residues in the two additive terms are complex conjugate of each other). The values of parameters in Eq. (\ref{propagator}) are taken from Ref. \cite{Maris} (with notations: 2CC stands for two pairs of complex conjugate
poles, 1R1CC stands for one pair of complex conjugate poles and one real pole, and 3R stands for three real poles, respectively). Just as was shown in Ref. \cite{Maris}, this propagator can be regarded as a good starting point for the study of low energy hadron physics because it can simultaneously describe the dynamical chiral symmetry breaking and quark confinement which are very important for low energy QCD. By means of this propagator, the authors 
in Refs. \cite{CON,HDK} calculated the pion mass and decay constant and the quark number susceptibility at finite chemical potential and zero temperature, respectively.

Using path integral methods, one can obtain the following well-known result
\begin{equation} 
\label{Pmurelation}
\rho(\mu)=\frac{\partial P(\mu)}{\partial \mu},
\end{equation}
where $P(\mu)$ is the pressure density. Integrating Eq. (\ref{Pmurelation}) gives
\begin{eqnarray}
\label{pressure1}
P(\mu)&=& P(\mu)|_{\mu=0}+\int_0^\mu d \mu' \rho(\mu') \nonumber \\
&=& P(\mu)|_{\mu=0}-N_c N_f Z_2 \int_0^\mu d \mu' \int \frac{d^4 p}{(2\pi)^4} tr\left[G_r(\mu')(p)\gamma_4\right] \nonumber \\
&=& P(\mu)|_{\mu=0}+\frac{2N_c N_f}{3\pi^2}\sum_{j=1}^{n_p}r_j
\theta\left(\mu'-\sqrt{\frac{\alpha_j+\sqrt{\alpha_j^2+\beta_j^2}}{2}}\,\,\right)
I(\mu;\alpha_j,\beta_j)
\end{eqnarray}
with
\begin{eqnarray}
\label{I}
I(\mu;\alpha_j,\beta_j)&&= \int_{\sqrt{\frac{\alpha_j+\sqrt{\alpha_j^2+\beta_j^2}}{2}}}^{\mu}~d\mu' \left(\mu'^2-\frac{\beta_j^2}{4\mu^2}-\alpha_j\right)^{3/2} {} \\
&&=\frac{3(\alpha_j^2-\beta_j^2)}{16}
\ln \frac{\sqrt{\mu^2-\alpha_j/2+\sqrt{\alpha_j^2+\beta_j^2}/2+
 \sqrt{\mu^2-\alpha_j/2-\sqrt{\alpha_j^2+\beta_j^2}/2}}}{\sqrt{\mu^2-\alpha_j/2+\sqrt{\alpha_j^2+\beta_j^2}/2}-
 \sqrt{\mu^2-\alpha_j/2-\sqrt{\alpha_j^2+\beta_j^2}/2}} {}\nonumber\\
&&+ \frac{3\alpha_j |\beta_j|}{4} \arctan \sqrt{\frac{(\sqrt{\alpha_j^2+\beta_j^2}-\alpha_j)
(\mu^2-\sqrt{\alpha_j^2+\beta_j^2}/2-\alpha_j/2)}{(\sqrt{\alpha_j^2+\beta_j^2}+\alpha_j)
(\mu^2+\sqrt{\alpha_j^2+\beta_j^2}/2-\alpha_j/2)}} {} \nonumber\\
&&+\frac{\mu^2}{4} \sqrt{\mu^4-\alpha_j
 \mu^2-\beta_j^2/4}-\frac{5 \alpha_j}{8}\sqrt{\mu^4-\alpha_j
 \mu^2-\beta_j^2/4}+\frac{\beta_j^2}{8}\frac{\sqrt{\mu^4-\alpha_j
 \mu^2-\beta_j^2/4}}{\mu^2}\nonumber,
\end{eqnarray}
where one has defined $\alpha_j +\beta_j i \equiv m_j^2 $, and in obtaining the last line of Eq. (\ref{pressure1}), one has made use of the quark propagator Eq. (\ref{propagator}) (for more details, please see Ref. \cite{zong2}). In Ref. \cite{zong2}, the constant term $P(\mu=0)$ (the pressure of the vacuum) in the EOS was omitted. This does not mean that the term $\mathcal{P}(\mu)|_{\mu=0}$ is unimportant. In fact, it is an important quantity. It enters the energy density, which is relevant for integrating the Tolman-Oppenheimer-Volkoff equations. At present it is not possible to calculate reliably $\mathcal{P}(\mu)|_{\mu=0}$  from first principles of QCD. In this paper, analogous to MIT bag model, we reconsider the effect of this term and think that there exists negative
pressure at zero chemical potential in vacuum which manifests the
confinement of QCD. Namely, we identify $P(\mu)|_{\mu=0}$ with $-B$, where $B$ is the vacuum bag constant. In addition, here, for simplicity, we do not
consider the chemical, thermodynamical equilibrium and electric
neutrality conditions, and assume that the star is made of pure three
flavor quark matter and have no crust structure. Meanwhile we
adopt the 2CC parameters. The energy density then reads
\begin{equation}
\label{eos}
 \varepsilon(\mu)=-P(\mu)+\mu\cdot\frac{\partial P}{\partial \mu}+B.
\end{equation}

\begin{figure}
\centerline{\epsfig{file=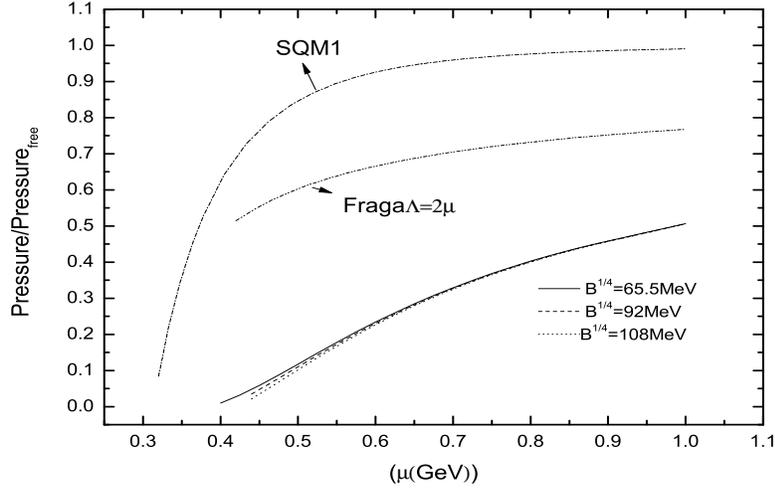,width=120mm,height=80mm} }
\vspace{-8mm}
\caption{The pressure as function of density, relative to the free
quark gas pressure $P_{free}=N_cN_f\mu^4/(12\pi^2)$. In the SQM1,
$B=(164.34~MeV)^4$
\label{fig:eos}}
\end{figure}

\begin{figure}
\centerline{\epsfig{file=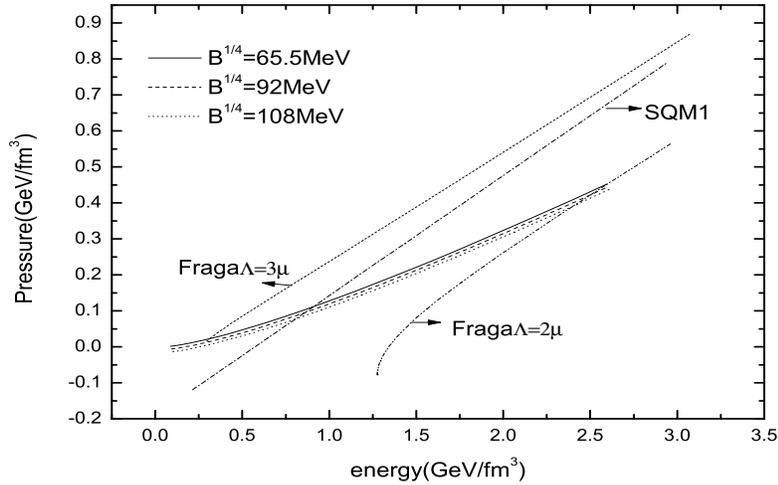,width=120mm,height=80mm} }
\vspace{-8mm}
\caption{Quark pressure-energy relation.
\label{fig:eos2}}
\end{figure}

\begin{figure}
\centerline{\epsfig{file=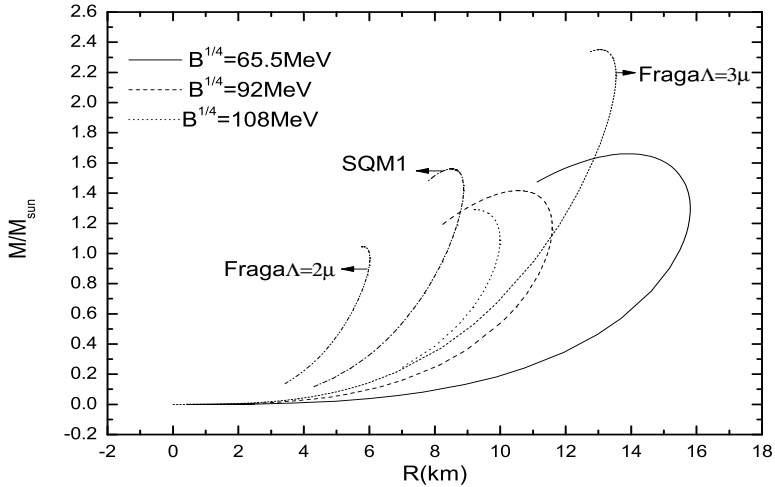,width=120mm,height=80mm}}
\vspace{-8mm}
\caption{Mass as function of Radius.
\label{fig:TOV1}}
\end{figure}

\begin{figure}
\centerline{\epsfig{file=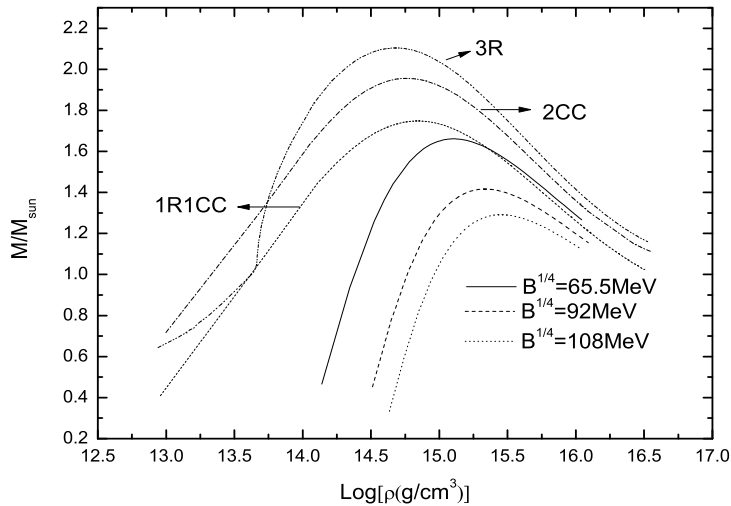,width=120mm,height=80mm} }
\vspace{-8mm}
\caption{Mass-Energy Density relation calculated with and without the presence of $B$\label{fig:TOV2}}
\end{figure}

In this work we take $B$ as an phenomenological model input and choose the following three values for $B$: $B=(65.5~MeV)^4$, $(92~MeV)^4$, $(108~MeV)^4$. Our EOS is plotted in Fig. \ref{fig:eos} and Fig. \ref{fig:eos2}, and analogous to Refs. \cite{Lattimer,Prakash}, we compare our results with other two models: Fraga {\it et~al.}'s perturbative QCD model and SQM1 model \cite{Baron}. From Fig. \ref{fig:eos} it can be seen that the pressures in our model for three different values of $B$ differ very little. It is clear that our EOS is softer than Fraga {\it et~al.}'s EOS ($\Lambda =2\mu$) and SQM1 model. In the SQM1 model, the value of $B$ is $(164.34~MeV)^4$ because of the energy limit of 939 MeV at zero pressure. To improve our EOS's softness, we have to adopt relatively small value of $B$ and take it as a phenomenological parameter only. While in Ref. \cite{Weber2}, the value of $B$ 
(namely, $(145~MeV)^4 < B < (162~MeV)^4 $) is determined by the requirement that the two flavor quark state is unstable though three flavor quark state is the true ground state. In Fig. \ref{fig:eos2} it can be seen more clearly that our EOS are relatively soft, Fraga {\it et~al.}'s EOS are moderate, while the SQM1 model is the stiffest. This can be easily explained: our EOS incorporates confinement and dynamical chiral symmetry breaking, so it is more appropriate to describe the property of quark matter at low density region. Fraga {\it et~al.}'s model is applicable to the massless quark gas in the chiral symmetry restoration region. The SQM1 model demonstrates the non-interacting relativistic free quark gas in the bag, so its EOS is the stiffest.
Applying this EOS, we obtain the structure of quark
star by integrating the Tolman-Oppenheimer-Volkoff (TOV) equation:
\begin{equation}
\label{TOV1}
\frac{dP(r)}{dr}=-\frac{G~(\varepsilon+P)(M+4 \pi r^3P)}{r(r-2GM)},
\end{equation}
\begin{equation}
\label{TOV2}
\frac{dM(r)}{dr}=4\pi r^2\varepsilon.
\end{equation}
The calculated results are shown in Fig. \ref{fig:TOV1} and we also
compare Fraga {\it et~al.}'s and SQM1's results with ours in this figure. For our model one can see that the scaling relation $M \sim B^{-1/2}$ is approximately consistent with Witten's result \cite{Witten}, meanwhile one notes that our radius of quark star are larger than that of Fraga {\it et~al.}'s model and that of SQM1 model. Then the density of star in our model is less than that in their models. In our results, according to three values of $B$, the maximum mass, radius and central number density are: mass 
$\sim$ 1.75 $M_\odot$, radius $\sim$ 14 Km, $Nc \sim 3.2~\rho_0$ for $B=(65.5~MeV)^4$, mass $\sim$1.42 $M_\odot$, radius $\sim$ 10.6 Km, $Nc \sim 5.2~\rho_0$ for $B=(92~MeV)^4$, mass$\sim$ 1.3 $M_\odot$, radius $\sim$ 9.2 Km, $Nc \sim 6.6~\rho_0$ for $B=(108~MeV)^4$, respectively. This phenomenon just reflects the fact that our EOS is appropriate to the low density region, while Fraga {\it et~al.}'s and SQM1's EOS work well at the high density area (namely, chiral symmetry restoration phase and asymptotic free region). The mass-energy density relations calculated with and without the presence of $B$ are illustrated in Fig. \ref{fig:TOV2}. The central density calculated with the presence of $B$ is obviously larger than the corresponding one calculated without the presence of $B$. This demonstrates that the star's compactness is promoted by adding the bag model constant. In this paper, we cannot consider the interior structure problem and assume that the star is homogeneous pure three flavors quark matter. Then there is not phase transition between the quark matter and the hadronic matter. But that is not imaginable. Actual stars do have inner structure (the crust structure has observable effects such as $\gamma$-burst, glitch and so on). Meanwhile we do not consider the chemical, thermodynamical equilibrium and electric neutrality conditions. So our model is only a rough approximation for the study of quark star. Here we should note that the main purpose of this work is to study the applicability of our EOS and qualitatively compare our results with those of other models. In the future work we should consider these factors.  

Although many models of quark star have been proposed in the literature, one may
still ask whether quark star exists, and if it exists, how to
distinguish it from neutron star. In Refs. \cite{Serot,Baym,Kisslinger},
the authors think that there is no quark matter core in the  neutron star. Particularly, in \"Ozel's article \cite{Ozel}, she concludes that it is impossible for quark core to exist in the central part of neutron star. Based on her observational data, she pointed out that the limit on the mass is $Mass~\geq 2.1~\pm 0.28~M_\odot$, the limit on the radius is $R~\geq 13.8~\pm 1.8 Km$. Therefore, those soft EOS will be ruled out. Of course, many people do not agree with her arguments (see, for example, Ref. \cite{Drago}). On the other hand, the authors in Ref. \cite{Kolb} think that $Cygnus X-3$ may be a strange quark star by measuring high energy cosmic ray from it. Usov argues that $\gamma$-ray bursters can be an observational
signal of strange quark star. By the pair creation mechanism of Coulomb barrier in Ref. \cite{Usov}, the computed result of emission energy coincides with the observational data. In cloudy bag models \cite{Thomas,Hosaka}, the pion decay may be the source of $\gamma$-ray burst and provides efficient cooling mechanism. The Kepler rotation period ($P_k$ may be lower than 1ms) is smaller than the neutron star with the same mass \cite{Weber3}. These can provide useful signal to distinguish quark star from neutron star. Nowadays we cannot draw a deterministic conclusion as to whether quark star exists, because we still do not have full knowledge about QCD and phase transition from nuclear matter to quark matter. 

To summarize, in this article, we present a new, renormalizable EOS at finite density and zero temperature and consider the effect of vacuum negative pressure. The pressures for different parameter $B$ in our model differ very little. The comparison between our EOS and the EOSs in other models show that our EOS is softer than other two models. The reason for the difference is discussed. Applying this EOS, one obtains the structure of quark star. The results are mass $\sim 1.75~M_\odot$, radius $\sim 14$ Km, $Nc \sim 3.2~\rho_0$ for $B=(65.5~MeV)^4$, mass $\sim 1.42 M_\odot$, radius $\sim 10.6$ Km, $Nc \sim 5.2~\rho_0$ for $B=(92~MeV)^4$, mass $\sim 1.3~M_\odot$, radius $\sim 9.2$ Km, $Nc \sim 6.6~\rho_0$ for $B=(108~MeV)^4$, respectively. By incorporating bag model, one can see that the compactness of our previous EOS is improved. 
\vspace{-7mm}
\begin{acknowledgments}
This work was supported in part by the National Natural Science Foundation of China (under Grant Nos 10775069 and 10935001) and the Research Fund for the Doctoral Program of Higher Education (under Grant No 200802840009).
\end{acknowledgments}


\vspace{-7mm}
\begin{thebibliography}{99}
\bibitem{shapiro}
A.L. Shapiro, S.A. Teukolsky, {\em Black Hole,White Dwarfs and Neutron Star: The Physics of Compact Objects} (New York, Wiley,1983).
\bibitem{Glendenning}
N.K. Glendenning, {\em Compact Stars --Nuclear Physics, Particle Physics, and General Relativity} (Springer, New York, 2000).
\bibitem{Gell-Mann}
M. Gell-Mann, Phys. Lett {\bf 8}, 214 (1964).
\bibitem{Zweig}
G. Zweig, Cern-Reports.~TH-401, TH-412 (1964).
\bibitem{Ivanenko}
D. Ivanenko and D.F. Kurdgelaidze, Lett. Nuov. Cement {\bf 2}, 13 (1969).
\bibitem{Itoh}
N. Itoh, Prog. Theor. Phys {\bf 44}, 291 (1970).
\bibitem{Iwamoto}
N. Iwamoto, Phys. Rev. Lett. {\bf 44}, 1637 (1980).
\bibitem{Bodmer}
A.R. Bodmer, Phys. Rev. D {\bf 4}, 1601 (1971).
\bibitem{Witten}
E. Witten, Phys. Rev. D {\bf 30}, 272 (1984).
\bibitem{Haensel}
 P. Haensel, J.L ~Zdunik and R. Schaeffer, Astron. Astrophys. {\bf 160}, 121 (1986).
\bibitem{Alcock}
C. Alcock, E. Farhi and A. Olinto, AP. J. {\bf 310} 261 (1986).
\bibitem{Weisskopf}
A. Chodos, R.L. Jaffe, K. Johnson, C.B. Thorn and V.F. Weisskopf, Phys. Rev. D {\bf 9}, 3471 (1974).
\bibitem{Weber}
F. Weber, {\em Pulse as Astrophysical Laboratories for Nuclear and Particle Physics} (Iop, Bristal, 1999).
\bibitem{Soff}
A. Peshier, B. Kampfer, and G. Soff, Phys. Rev. C {\bf 61}, 045203 (2000).
\bibitem{Paris}
M. Alford, M. Braby, M. Paris, and S. Reddy, AP. J. {\bf 629}, 969 (2005).
\bibitem{Rehberg}
P. Rehberg, S.P. Klevansky, and J. H\"ufner, Phys. Rev. C  {\bf 53}, 410(1996).
\bibitem{Greiner}
M. Hanauske, L.M. Satarov, I.N. Mishustin, H. Stocker, and W. Greiner, Phys. Rev. D {\bf 64}, 043005 (2001).
\bibitem{Ruster}
S.B. R\"uster, D.H. Rischke, Phys. Rev. D {\bf 69}, 045011 (2004).
\bibitem{Menezes}
D.P. Menezes, C. Providencia, D.B. Melrose, J. Phys. G: Nucl. Part. Phys, {\bf 32},1081 (2006).
\bibitem{Freedman}
B. Freedman and L.McLerran, Phys. Rev. D {\bf 16}, 1130 (1977); ${\it ibid.}$ {\bf 16}, 1147 (1977); ${\it ibid.}$ {\bf 16}, 1169 (1977); ${\it ibid.}$ {\bf 17}, 1109 (1978).
\bibitem{Baluni}
V. Baluni, Phys. Rev. D {\bf 17}, 2092 (1978).
\bibitem{Fraga}
E.S. Fraga, R.D. Pisarski and J. Schaffner-Bielich, Phys. Rev.D {\bf 63}, 121702 (2001).
\bibitem{Farhi}
E. Farhi and R.L. Jaffe, Phys. Rev. D {\bf 30}, 2379 (1984).
\bibitem{Thomas}
A.W. Thomas, Czech. J. Phys. {\bf 32}, 239 (1982).
\bibitem{Hosaka}
A. Hosaka and H. Toki, Phys. Rept. {\bf 277}, 65 (1996).
\bibitem{Klevansky}
S.P. Klevansky, Review of Modern Physics, {\bf 64}, 649 (1992).
\bibitem{Kapusta}
J.I. Kapusta, {\em Finite Temperature Field Theory } (Cambridge University Press, Cambridge,1989).
\bibitem{Sun}
W.M. Sun and H.S. Zong, J. Mod. Phys. A {\bf 22}, 3201 (2007).
\bibitem{Roberts}
C.D. Roberts and A.G. Williams, Prog. Part. Nucl. Phys {\bf 33} ,477 (1994).
\bibitem{Schmidt}
C.D. Roberts and S.M. Schmidt, Prog. Part. Nucl. Phys {\bf 45 S1}, 1 (2000).
\bibitem{Roberts1}
P. Maris and C.D. Roberts, Int. J. Mod. Phys. E {\bf 12 }, 297 (2003).
\bibitem{Zong}
H.S. Zong, L. Chang, F.Y. Hou, W.M. Sun and Y.X. Liu, Phys. Rev. C {\bf 71}, 015205 (2005).
\bibitem{Feng}
H.T. Feng, F.Y. Hou, X. He, W.M. Sun, and H.S. Zong, Phys. Rev. D {\bf 73}, 016004 (2006).
\bibitem{He}
H.T. Feng, W.M. Sun, D.K. He, and H.S. Zong, Phys. Lett. B {\bf 661} 57 (2008).
\bibitem{XY} X.Y. Li, X.F. L$\ddot{u}$, B. Wang, W.M. Sun, and H.S. Zong, Phys. Rev. C {\bf 80}, 034909 (2009).
\bibitem{Y} Y. Jiang, H. Li, S.S. Huang, W.M. Sun, and H.S. Zong, J. Phys. G {\bf 37}, 105004 (2010).
\bibitem{Maris} R. Alkofer, W. Detmold, C.S. Fischer and P. Maris, Phys. Rev. D {\bf 70}, 014014 (2004).
\bibitem{zong2}
H.S. Zong, and W.M. Sun, Phys. Rev. D {\bf 78}, 054001 (2008).
\bibitem{Tandy} 
M. S. Bhagwat, M. A. Pichowsky, and P. C. Tandy, Phys. Rev. D {\bf 67}, 054019 (2003).
\bibitem{CON} 
Y. Jiang, Y. B. Zhang, W. M. Sun, and H. S. Zong, Phys. Rev. D {\bf 78}, 014005 (2008).
\bibitem{HDK} 
D. K. He, X. X. Ruan, Y. Jiang, W. M. Sun, and H. S. Zong,  Phys. Lett. B {\bf 680}, 432 (2009).
\bibitem{Lattimer}
J.M. Lattimer and M. Prakash, AP. J. {\bf 550} 426 (2001).
\bibitem{Prakash}
J.M. Lattimer and M. Prakash, Science. {\bf 304}, 536 (2004).
\bibitem{Baron}
M. Prakash, E. Baron, and M. Prakash, Phys. Lett. B {\bf 243} 175 (1990).
\bibitem{Weber2}
F. Weber, Prog. Part. Nucl. Phys. {\bf54}, 193 (2005).
\bibitem{Serot}
B.D. Serot, H. Uechi, Ann. Phys, {\bf 179}, 272 (1987).
\bibitem{Baym}
G. Baym, S. A. Chin, Phys. Lett. B {\bf 62}, 241 (1975).
\bibitem{Kisslinger}
B.D. Keister, L.S. Kisslinger, Phys.\ Lett.\ B {\bf 64}, 117 (1976).
\bibitem{Ozel}
F. \"Ozel, Nature {\bf 441}, 1115 (2006).
\bibitem{Drago}
M. Alford, D. Blaschke, A. Drago, T. Kl\"ahn, G. Pagliara, J. Schaffner-Bielich, Nature {\bf 445}, E7 (2007).
\bibitem{Kolb}
G. Baym, E. Kolb, L. Mclerran, and T.P. Walker, Phys. Lett. B {\bf 160} 181 (1985).
\bibitem{Usov}
V.V. Usov, Phys. Rev. Lett. {\bf 80}, 230 (1998).
\bibitem{Weber3}
F. Weber, Pulsars as Astrophysical Laboratoriers for nuclear and
particle physics, High Energy Physics,Cosmology and Gravitation
Series (IOP publishing, Bristol,Great Britain, 1999)

\end{thebibliography}
\end{document}